\documentclass[useAMS,usenatbib,a4paper,referee]{mn2e}
\usepackage{graphicx}
\usepackage{amsmath}
\usepackage{txfonts}
\usepackage{mathrsfs}

\def\aap{A\&A}
\def\apjl{ApJ}

\def\apjs{ApJS}

\newcommand{\be}{\begin{equation}}
\newcommand{\ee}{\end{equation}}
\newcommand{\bea}{\begin{eqnarray}}
\newcommand{\eea}{\end{eqnarray}}

\title[Redshift Drift]{Definitive Test of the $R_{\rm h}=ct$ Universe Using Redshift Drift}
\author[Fulvio Melia]{Fulvio Melia\thanks{John Woodruff Simpson 
Fellow. E-mail: fmelia@email.arizona.edu} \\
\null Department of Physics, The Applied Math Program, and Department of Astronomy, 
The University of Arizona, AZ 85721, USA}

\begin{document}

\date{}

\pagerange{\pageref{firstpage}--\pageref{lastpage}} \pubyear{2013}

\maketitle

\label{firstpage}

\begin{abstract}
The redshift drift of objects moving in the Hubble flow has been proposed as a powerful 
model-independent probe of the underlying cosmology. A measurement of the first and second 
order redshift derivatives appears to be well within the reach of upcoming surveys 
using ELT-HIRES and the SKA Phase 2 array. Here we show that an unambiguous prediction of 
the $R_{\rm h}=ct$ cosmology is {\it zero} drift at all redshifts, contrasting sharply with 
all other models in which the expansion rate is variable. For example, multi-year monitoring
of sources at redshift $z=5$ with the ELT-HIRES is expected to show a velocity
shift $\Delta v = -15$ cm s$^{-1}$ yr$^{-1}$ due to the redshift drift in {\it Planck}
$\Lambda$CDM, while $\Delta v=0$ cm s$^{-1}$ yr$^{-1}$ in $R_{\rm h}=ct$. With an 
anticipated ELT-HIRES measurement error of $\pm 5$ cm s$^{-1}$ yr$^{-1}$ after 5 years, 
these upcoming redshift drift measurements might therefore be able to differentiate 
between $R_{\rm h}=ct$ and {\it Planck} $\Lambda$CDM at $\sim 3\sigma$, assuming
that any possible source evolution is well understood. Such a result would provide the 
strongest evidence yet in favour of the $R_{\rm h}=ct$ cosmology. With a $20$-year 
baseline, these observations could favor one of these models over the other at better 
than $5\sigma$.
\end{abstract}


\section{Introduction} 
The cosmological spacetime is now being studied using several techniques,
including the use of Type~Ia SNe as standard candles (Riess et~al.\ 1998; Perlmutter
et~al.\ 1999), baryon acoustic oscillations (Seo \& Eisenstein 2003; Eisenstein 
et~al.\ 2005; Pritchard et~al.\ 2007; Percival et~al.\ 2007), weak lensing (Refregier 
2003), and also cluster counts (Haiman et~al.\ 2000); the range of observations is
actually much larger than this. Most of these methods, however, depend on integrated 
quantities, such as the angular or luminosity distances, that depend on the
assumed cosmology.  Such data cannot easily be used for unbiased
comparisons of different expansion histories.  For example, in the case
of Type~Ia SNe, at least three or four `nuisance' parameters characterizing 
the SN lightcurve must be optimized along with the model's free parameters, 
making the data compliant to the underlying cosmology (Melia 2012; Wei 
et al. 2015).

Nonetheless, recent progress comparing predictions of the $R_{\rm h}=ct$ 
Universe (Melia 2007; Melia \& Shevchuk 2012; Melia 2016a, 2016b) and $\Lambda$CDM 
versus the observations suggests that the $R_{\rm h}=ct$ cosmology may
be a better fit to the data, both at high redshifts where, e.g., the angular correlation
function in the cosmic microwave background (CMB) disfavors the standard
model (Melia 2014a) and the premature appearance of galaxies (Melia 2014b) and 
supermassive quasars (Melia 2013a; Melia \& McClintock 2015) may be more easily 
explained by $R_{\rm h}=ct$ than $\Lambda$CDM, and at low and intermediate 
redshifts, as seen, e.g., in the Type Ia SN Hubble diagram (Wei et al. 2015) and 
the distribution of baryon acoustic oscillations  (BAO; Melia \& L\'opez-Corredoira
2016 and references cited therein). 

The Alcock-Paczy\'nski test based on these BAO measurements is particularly 
noteworthy because it does not depend on any possible source evolution, and is 
strictly based on the geometry of the Universe through the changing ratio of 
angular to spatial/redshift size of spherically-symmetric distributions with 
distance. With $\sim 4\%$ accuracy now available for the positioning of the 
BAO peak, the latest measurements appear to rule out the standard model at
better than $\sim 2.6\,\sigma$ (Melia \& L\'opez-Corredoira 2016). This in
itself is rather important, but is even more significant because, in contrast
to $\Lambda$CDM, these same data suggest that the probability of $R_{\rm h}=ct$ 
being correct is close to unity. 

Many of these model comparisons, however, are not yet conclusive when
the astrophysical processes underlying the behavior of the sources are still not fully
understood. For example, in the case of the premature appearance of supermassive
black holes (at redshifts $\sim 6-7$), we may simply be dealing with the creation
of massive seeds ($\sim 10^5\;M_\odot$) in the early Universe, or episodic
super-Eddington accretion, rendering their timeline consistent with the
predictions of the standard model. Similarly, the early appearance of galaxies
(at $z\sim 10-12$) may itself be due to anomalously high star-formation
rates during a period, e.g., the transition from Population III to Population II
stars, that is still in need of further theoretical refinement.

So we now find ourselves in a situation where some tension is emerging
between the predictions of the standard cosmology and the high precision
measurements (e.g., the Alcock-Paczy\'nski effect based on BAO peak
localization), which suggests that either the underlying cosmological
model needs to evolve---perhaps in the direction of $R_{\rm h}=ct$---or
that we need to refine our understanding of the physics responsible
for the behavior of the sources we use for these measurements. There
is therefore ample motivation for finding new, even better and more
precise means of comparing different models.

The purpose of this letter is to demonstrate that a measurement of the
redshift drift over a baseline of 5-20 years should provide the best test
yet for differentiating between $\Lambda$CDM and $R_{\rm h}=ct$, 
with an eventual preference of one cosmology over the other at a 
confidence level approaching $3-5\sigma$. 

\section{Redshift Drift}

The redshift drift of objects in the Hubble flow is a direct non-geometric 
probe of the dynamics of the Universe that does not rely on assumptions
concerning gravity and clustering (Sandage 1962). It merely requires the 
validity of the Cosmological principle, asserting that the Universe is 
homogeneous and isotropic on large scales. For a fixed comoving distance,
the redshift of a source changes with time in a Universe with a variable
expansion rate, so its first and second derivatives may be used to
distinguish between different models (Corasaniti et al. 2007; Quercellini
et al. 2012; Martins et al. 2016). High-resolution spectrographs (Loeb 1998;
Liske et al. 2008), such as ELT-HIRES (Liske et al. 2014), will allow measurements
in the approximate redshift range $2\lesssim z\lesssim 5$. Below $z\sim 1$, 
measurements may be made with the SKA Phase 2 array (Klockner et al. 2015),
and possibly also with 21cm experiments, such as CHIME (Yu et al. 2014).

The relevant equations for the first (and second) time derivatives of the
redshift in the context of $\Lambda$CDM have been derived by, e.g., Weinberg
(1972), Liske et al. (2008) and Martins et al. (2016), and we here simply 
quote their key results. Defining the cosmological redshift, $z$, between 
the time of emission, $t_e$, when the expansion factor $a(t)$ had the value 
$a_e$, and the time of observation, $t_0$, when $a(t)=a_0$, as
\begin{equation}
1+z={a_0\over a_e}\;,
\end{equation}
we may write its derivative with respect to the current cosmic time as
\begin{equation}
{dz\over dt_0} = [1+z(t_0)]H(t_0)-{a_0\over a_e^2}{da_e\over dt_e}{dt_e\over dt_0}
\end{equation}
where, by definition,
\begin{equation}
H(t) = {1\over a(t)}{da(t)\over dt}
\end{equation}
is the Hubble constant at time $t$. But
\begin{equation}
dt_0=[1+z(t_0)]\,dt_e\;,
\end{equation}
so
\begin{equation}
{dz\over dt_0}=[1+z]H_0-H(z)\;.
\end{equation}
Here we may write 
\begin{equation}
H(z) = H_0\,E(z)\;,
\end{equation}
where $H_0\equiv H(t_0)$ is the Hubble constant today, and
\begin{equation}
E^2(z)=\Omega_m(1+z)^3+\Omega_r(1+z)^4+\Omega_{de}(1+z)^{3(1+w_{de})}+\Omega_k(1+z)^2\;.
\end{equation}
In this expression, $\Omega_m$, $\Omega_r$ and $\Omega_{de}$ are the ratios of energy 
density $\rho_m$ (matter), $\rho_r$ (radiation) and $\rho_{de}$ (dark energy) to the 
current critical density $\rho_c\equiv 3c^2H_0^2/8\pi G$, and $w_{de}$ is the dark-energy
equation of state parameter, $w_{de}\equiv p_{de}/\rho_{de}$. In addition, the ratio 
$\Omega_k\equiv -kc^2/(a_0^2\,\rho_c)$ is non-zero when the Universe is spatially 
curved, i.e., when $k\not=0$. 

During a monitoring campaign, the surveys measure the spectroscopic velocity
shift $\Delta v$ associated with the redshift drift $\Delta z$ over a time interval 
$\Delta t$. These two quantities are related via the expression 
\begin{equation}
\Delta v={c\Delta z\over 1+z}={c\Delta t\over 1+z}{dz\over dt_0}\;.
\end{equation}
In figure~1, we plot $\Delta v$ as a function of $z$ for the {\it Planck} $\Lambda$CDM 
model (defined by the parameter values $k=0$, $\Omega_m=0.3$, $H_0=67.8$ km s$^{-1}$
Mpc$^{-1}$ and $w_{de}=-1$; solid black curve), and a slight variation (short dash black 
curve) with $\Omega_m=0.28$ to illustrate the change expected with redshift and 
alternative values of the parameters. 

\begin{figure}
\center{\includegraphics[scale=1.1,angle=0]{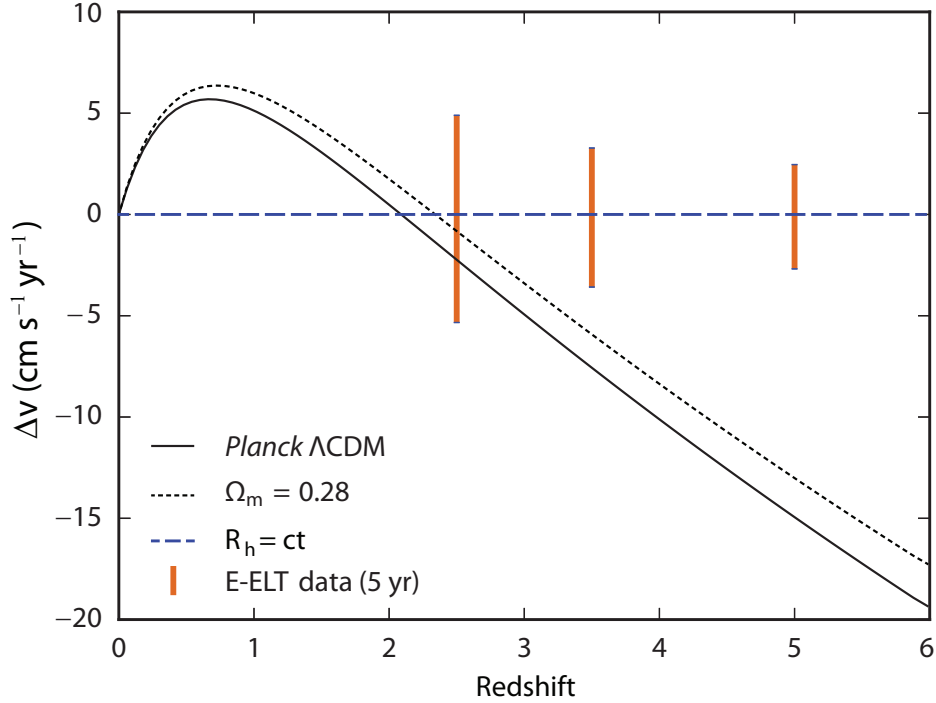}
\caption{Expected velocity shift $\Delta v$ associated with the redshift drift for the
{\it Planck} $\Lambda$CDM model ($k=0$, $\Omega_m=0.3$ $H_0=67.8$ km s$^{-1}$
Mpc$^{-1}$; solid black), a slight variation with $\Omega_m=0.28$, and the $R_{\rm h}=ct$
universe (blue long dash), in which $\Delta v=dz/dt_0=0$ at all redshifts. Also shown are
the expected $1\sigma$ errors (red) at $z=2.5$, $3.5$, and $5.0$ with the ELT-HIRES after 
5 years of monitoring. (Adapted from Martins et al. 2016)}}
\end{figure}

In $R_{\rm h}=ct$, the situation is much simpler. Since $a(t)=t/t_0$ in this Universe
(see, e.g., Melia 2007, 2016a, 2016b; Melia \& Shevchuk 2012), we have
\begin{equation}
H(t) \equiv {\dot{a}\over a}={1\over t}\;.
\end{equation}
But from Equation~(1), we also see that $(1+z)=t_0/t_e$. Therefore, 
$(1+z)=H(t_e)/H(t_0)$, or
\begin{equation}
H(t_e)=H(t_0)[1+z(t_0)]\;.
\end{equation}
Thus, we conclude from Equation~(5) that 
\begin{equation}
{dz\over dt_0}=0\;,
\end{equation}
and therefore
\begin{equation}
\Delta v =0
\end{equation}
at all redshifts, shown as the blue long-dashed line in figure~1.

According to Liske et al. (2008), the ELT-HIRES is expected to observe the
spectroscopic velocity shift with an uncertainty of
\begin{equation}
\sigma_{\Delta v} = 1.35 {2370\over {\rm S/N}}\sqrt{{30\over N_{\rm QSO}}}\left({5
\over 1+z_{\rm QSO}}\right)^\alpha\;,
\end{equation}
where $\alpha=1.7$ for $z\le 4$, and $\alpha=0.9$ for $z>4$, with a ${\rm S/N}$
of approximately 1,500 after 5 years of monitoring $N_{\rm QSO}=10$ quasars
in each of three redshift bins at $z=2.5$, $3.5$ and $5.0$. These uncertainties
(shown in red) are approximately $12$, $8$ and $5$ cm s$^{-1}$ yr$^{-1}$ at
these redshifts. With a baseline of 20 years, the uncertaintes are reduced to
approximately $6$, $4$ and $3$ cm s$^{-1}$ yr$^{-1}$, respectively. Thus, we 
may already see a $\sim 3\sigma$ difference between the predicted velocity
shifts in these two models at $z=5$ after only 5 years. The difference increases
to $\sim 5\sigma$ after 20 years.

\section{Conclusion}
Unlike many other types of cosmological probes, a measurement of the redshift
drift of sources moving passively with the Hubble flow offers us the possibility
of watching the Universe expand in real time. It does not rely on integrated
quantities, such as the luminosity distance, and therefore does not require
the pre-assumption of any particular model. Over the next few years, it will
be possible to monitor distant sources spectroscopically in order to measure
the velocity shift associated with this redshift drift. An accessible goal
of this work ought to be a direct one-on-one comparison between the $R_{\rm h}=ct$
Universe and {\it Planck} $\Lambda$CDM. The former firmly predicts zero redshift
drift at all redshifts, easily distinguishable from all other models associated
with a variable expansion rate. Over a 20-year baseline, these measurements will
strongly favour one of these models over the other at a $\sim 5\sigma$ confidence
level. However, the velocity shifts predicted by these two models at
$z\sim 5$ are so different ($-15$ cm s$^{-1}$ yr$^{-1}$ for the former
versus $0$ cm s$^{-1}$ yr$^{-1}$ for the latter), that even after only 
5 years the expected $\sigma_{\Delta v}\sim 5$ cm s$^{-1}$ yr$^{-1}$ accuracy 
of the ELT-HIRES may already be sufficient to rule out one of these models
relative to the other at a confidence level approaching $3\sigma$.

Should $R_{\rm h}=ct$ emerge as the correct cosmology, the consequences are,
of course, quite profound. The reason is that, although certain observational
signatures, such as the luminosity distance, can be somewhat similar for these
two models at low redshifts (see figure 3 in Melia 2015), they diverge rather
quickly in the early universe. So much so that while inflation is required
to solve the horizon problem in $\Lambda$CDM, it is not needed (and probably
never happend) in $R_{\rm h}=ct$ (Melia 2013b).

\section*{Acknowledgments}
I am grateful to Ignacio Trujillo for suggesting this test, and to the referee, Pier
Stefano Corasaniti, for helpful suggestions that have led to an improvement
in the presentation of the manuscript. I also acknowledge Amherst College 
for its support through a John Woodruff Simpson Fellowship. Part of this work 
was carried out at the Instituto de Astrofísica de Canarias in Tenerife.

\end{document}